\documentclass[10pt]{article}
\usepackage{latexsym}
\usepackage{amssymb}
\usepackage{amsmath}
\usepackage{amscd}
\usepackage{amsthm}
\usepackage{epsfig}
\usepackage[left=2.5cm,top=2.5cm,right=2.0cm,bottom=2.0cm]{geometry}
\usepackage{hyperref}
\usepackage{textcomp}
\begin{document}
\begin{center}
\Large{\bf{Similarity dark energy models in Bianchi type - I space-time}}

\vspace{5mm}

\normalsize{Ahmad T Ali$^{\dag,\ddag}$}, Anil Kumar Yadav$^{\S}$ and Abdulah K Alzahrani$^{\dag}$\\

\vspace{2mm} \normalsize{$^\dag$ King Abdul Aziz University,
Faculty of Science, Department of Mathematics,\\
PO Box 80203, Jeddah, 21589, Saudi Arabia.}\\
E-mail: ahmadtawfik95@gmail.com\\
\normalsize{$^\ddag$ Mathematics Department, Faculty of Science, Al-Azhar University,\\
Nasr city, 11884, Cairo, Egypt.}\\
\vspace{2mm} \normalsize{$^\S$ Department of Physics, United College of Engineering $\&$ Research,\\
 Greater Noida - 201306, India.\\
E-mail: abanilyadav@yahoo.co.in}\\
\end{center}

%%%%%%%%%%%%%%%%%%%%%%%%%%%%%%%%%%%%%%%%%%%%%%%%%%%%%%%%%%
%\date{}
\begin{abstract}

We investigate some new similarity inhomogeneous solutions of anisotropic
dark energy and perfect fluid in Bianchi type-I space-time. Three
different time dependent skewness parameters along the spatial
directions are introduced to quantify the deviation of pressure from
isotropy.  We consider the case when the dark energy is minimally
coupled to the perfect fluid as well as direct interaction with it.
The Lie symmetry generators that leave the equation invariant are
identified and we generate an optimal system of one-dimensional subalgebras.
Each element of the optimal system is used to reduce the partial differential
equation to an ordinary differential equation which is further analyzed.
We solve the Einstein field equations, described by a system of non-linear
partial differential equations (NLPDEs), by using the Lie point symmetry
analysis method. The geometrical and kinematical features of the
models and the behavior of the anisotropy of dark energy, are
examined in detail.

\end{abstract}



%%%%%%%%%%%%%%%%%%%%%%%%%%%%%%%%%%%%%%%%%%%%%%%
\section{Introduction }
%%%%%%%%%%%%%%%%%%%%%%%%%%%%%%%%%%%%%%%%%%%%%%%
The overall energy budget of the universe is dictated by dark matter and dark energy with
a minor contamination from baryonic matter, where dark energy is supposed to dominate
the cosmic landscape causing the acceleration of universe. Since 1998,
we have witnessed astrophysical observations, still we are struggling to find the suitable
candidates for dark energy from fundamental physics. Many cosmologists believe that
the simplest candidate for the dark energy is the cosmological constant $(\Lambda)$ since it fits well
with observational data. During the cosmological evolution, the $\Lambda$-term has the density density
and pressure $p^{(de)} = -\rho^{(de)}$. However, one has the reason to
dislike $\Lambda$-term because it always suffers from the "fine-tunning" problems and ``cosmic
coincidence'' puzzles \cite{cope2006} on the theoretical ground.\\

The WMAP observations and Plank data had been favored the asymmetric expansion of universe and showing anomalies in Large scale structure \cite{hinshaw2009, watanabe2009, ade2014}. Few years ago, Companelli et al \cite{campanelli2007,campanelli2009} and Jaffe et al \cite{jaffe2006} had investigated some anisotropic models in consistency with observations \cite{hinshaw2009, watanabe2009}. Bianchi - I model is the simplest model among the all anisotropic models and it give an opportunity of non-symmetrical expansion of matter/energy along $x$, $y$ and $z$ directions. Some applications of Bianchi - I madels in cosmology have been elaborated in refs. \cite{yadav2012,yadav2016} and \cite{yadav2011}$-$\cite{mishra2015}. Lie group of transformations have been extensively applied to linear and nonlinear differential equations in 
the area of theoretical physics such as: general relativity, particle physics and cosmology \cite{baum1, ibra1}. 
The method of Lie symmetry group is one of the most useful tools for finding exact solutions for the 
Einstein field equations described by a system of NLPDEs. \cite{blum1, blum2, olve1, ovsi1, step1}. 
Recently we have developed a formalism to solve non linear Einstein's field equations in general relativity 
\cite{ali2014, ali2014a, anil2014, ali2014b}.\\

In our earlier work \cite{anil2014}, we have proposed the invariant solution of 
dark energy (DE) model in cylindrically symmetric space-time while in this paper, we confine ourselves to investigate the 
similarity solution of DE model in Bianchi - I space-time which is entirely different from \cite{anil2014}. 
In the Astrophysical community, the 
inhomogeneous cosmological models have gained interest due to exact perturbation of FRW model and are more often 
employed to study cosmological phenomenon. That is why, here, we consider inhomogeneous Bianchi - I space-time. 
The paper is organized as follows: In section 2, the basic formalism for anisotropic DE has been
discussed for an anisotropic and spatially homogeneous Bianchi I space-time. Similar formalism has been already developed
in our earlier work \cite{anil2014}. Section 3 deals with the Lie point symmetry and similarity solutions of the Bianchi I models. The Physical aspects of similarity dark energy models are given in section 4.
Finally the conclusions are summarized in section 5.

%%%%%%%%%%%%%%%%%%%%%%%%%%%%%%%%%%%%%%%%%%%%%%%
\section{The metric and field equations}
%%%%%%%%%%%%%%%%%%%%%%%%%%%%%%%%%%%%%%%%%%%%%%%
The Bianchi type-I space-time is given by
\begin{equation}
 \label{spacetime}
ds^{2}=A^{2}\,dx^{2}+B^{2}\,dy^{2}+C^2\,dz^{2}-dt^{2},
\end{equation}
where the metric potentials $A$, $B$ and $C$ are functions of $x$
and $t$. Einstien's field equations in the case of a mixture of
perfect fluid and anisotropic dark energy are given by
\begin{equation}
 \label{efe}
G^{i}_{j}=R^{i}_{j}-\frac{1}{2}g^{i}_{j}=-T^{(pf)i}_{j}-T^{(de)i}_{j},
\end{equation}
with
\begin{equation}
 \label{pf}
T^{(pf)i}_{j} = diag[-\rho^{(pf)}, p^{(pf)}, p^{(pf)},
p^{(pf)}]=diag[-1, \omega^{(pf)}, \omega^{(pf)},
\omega^{(pf)}]\rho^{(pf)}
\end{equation}
and
\begin{equation}
 \label{de}
T^{(de)i}_{j}=diag[-\rho^{(de)}, p^{(de)}_{x}, p^{(de)}_{y},
p^{(de)}_{z}] =diag[-1, \omega_x^{(de)}, \omega_y^{(de)},
\omega_z^{(de)}]\rho^{(de)} = diag[-1, \omega+\delta, \omega+\gamma,
\omega+\eta]\rho^{(de)}
\end{equation}
where $g^{i}_{j}$ are the metric tensor with $g_{ij}u^iu^j=-1$;
$u^i$ is the flow vector; $R^{i}_{j}$ is the Ricci tensor;
$R=R^{i}_{i}$ is the Ricci scalar; $p^{(pf)}$, $\rho^{(pf)}$ and
$\rho^{(de)}$ are, respectively the pressure and energy density of
the perfect fluid and dark energy components; $\omega^{(pf)}$ is the
EoS parameter of perfect fluid with $\omega^{(pf)}\geq0$;
$\omega_x^{(de)}$, $\omega_y^{(de)}$ and $\omega_z^{(de)}$ are the
deviation-free EoS parameters of dark energy, respectively, on the
$x$, $y$ and $z$ axis; $\delta$, $\gamma$ and $\eta$ are skewness
parameters along $x$, $y$ and $z$ axis respectively, which modify
equation of state parameter of dark energy. Here, $\delta$, $\gamma$
and $\eta$ are not necessarily constants and can be functions of the
cosmic time $t$.

In co-moving coordinate system, the field equation (\ref{efe}), for the inhomogeneous
space-time (\ref{spacetime}), read as

\begin{equation}
 \label{efe1}
p^{(pf)}+\omega_x^{(de)}\rho^{(de)}=\frac{B'C'}{A^{2}BC}-
\frac{\dot{B}\dot{C}}{BC}-\frac{\ddot{B}}{B}-\frac{\ddot{C}}{C},
\end{equation}

\begin{equation}
 \label{efe2}
p^{(pf)}+\omega_y^{(de)}\rho^{(de)}=\frac{1}{A^{2}}\left[\frac{C''}{C}-
\frac{A'C'}{AC}\right]-\frac{\ddot{A}}{A}-\frac{\dot{A}\dot{C}}{AC}-\frac{\ddot{C}}{C},
\end{equation}

\begin{equation}
 \label{efe3}
p^{(pf)}+\omega_z^{(de)}\rho^{(de)}=\frac{1}{A^{2}}\left[\frac{B''}{B}-
\frac{A'B'}{AB}\right]-\frac{\ddot{A}}{A}-\frac{\dot{A}\dot{B}}{AB}-\frac{\ddot{B}}{B},
\end{equation}

\begin{equation}
 \label{efe4}
\rho^{(pf)}+\rho^{(de)}=\frac{1}{A^{2}}\left[\frac{B''}{B}+
\frac{B'C'}{BC}+\frac{C''}{C}-\frac{A'B'}{AB}-\frac{A'C'}{AC}\right]-\frac{\dot{A}\dot{B}}{AB}-\frac{\dot{A}\dot{C}}{AC}-\frac{\dot{B}\dot{C}}{BC},
\end{equation}

\begin{equation}
 \label{efe5}
\frac{\dot{C}^{\prime}}{C}+\frac{\dot{B}^{\prime}}{B}=\frac{\dot{A}}{A}\Big[\frac{C^{\prime}}{C}+\frac{B^{\prime}}{B}\Big].
\end{equation}
Here $A^{\prime}=\frac{dA}{dx}$, $\dot{A} = \frac{dA}{dt}$ and so on.\\

The velocity field $u^i$ is ir-rotational. The scalar expansion
$\Theta$, shear scalar $\sigma^2$, acceleration vector $\dot{u}_i$
and proper volume $V$ are respectively found from the following
expressions \cite{fein1, rayc1}:

\begin{equation}  \label{u215}
\Theta\,=\,u_{;i}^{i}=\dfrac{\dot{A}}{A}+\dfrac{\dot{B}}{B}+\dfrac{\dot{C}}{C},
\end{equation}

\begin{equation}  \label{u216}
%\left\{
\begin{array}{ll}
\sigma^2\,=\,\dfrac{1}{2}\,\sigma_{ij}\,\sigma^{ij}=\dfrac{\Theta^2}{3}-\dfrac{\dot{A}\dot{B}}{AB}-\dfrac{\dot{A}\dot{C}}{AC}-\dfrac{\dot{B}\dot{C}}{BC},
\end{array}
%\right.
\end{equation}

\begin{equation}  \label{u217}
\dot{u}_i\,=\,u_{i;j}\,u^j\,=\,\big(0,0,0,0\big),
\end{equation}

\begin{equation}  \label{u218}
V=\sqrt{-g}=A\,B\,C,
\end{equation}
where $g$ is the determinant of the metric (\ref{spacetime}). The shear tensor is
\begin{equation}  \label{u219}
  \begin{array}{ll}
\sigma_{ij}\,=\,u_{(i;j)}+\dot{u}_{(i}\,u_{j)}-\frac{1}{3}\,\Theta\,(g_{ij}+u_i\,u_j).
\end{array}
\end{equation}
and the non-vanishing components of the $\sigma_i^j$ are
\begin{equation}  \label{u220}
\left\{
  \begin{array}{ll}
    \sigma_1^1\,&=\,\dfrac{1}{3}\Big(\dfrac{2\dot{A}}{A}-\dfrac{\dot{B}}{B}-\dfrac{\dot{C}}{C}\Big),\,\,\,\,\,\,\,\,\,\,
\sigma_2^2\,=\,\dfrac{1}{3}\Big(\dfrac{2\dot{B}}{B}-\dfrac{\dot{A}}{A}-\dfrac{\dot{C}}{C}\Big),\\
\\
\sigma_3^3\,&=\,\dfrac{1}{3}\Big(\dfrac{2\dot{C}}{C}-\dfrac{\dot{B}}{B}-\dfrac{\dot{A}}{A}\Big),\,\,\,\,\,\,\,\,\,\,
\sigma_4^4\,=0.
   \end{array}
\right.
\end{equation}

The Einstein's field equations (\ref{efe1})-(\ref{efe5}) constitute
a system of five highly NLPDEs with six unknowns variables, $A$,
$B$, $C$, $p^{(pf)}$, $\rho^{(pf)}$ and $\rho^{(de)}$. Therefore,
one physically reasonable conditions amongst these parameters are
required to obtain explicit solutions of the field equations. Let us
assume that the metric potential function $A$ is a function of the
time only, i.e., $A(x,t)=A(t)$. If we substitute the metric function
$A(x,t)=A(t)$ in the Einstein field equations, the equations
(\ref{efe1})-(\ref{efe5}) transform to the NLPDEs of the
coefficients $B$ and $C$  only, as the following new form:
\begin{equation}  \label{u210-1}
  %\left{
\begin{array}{ll}
    E_1=\dfrac{\dot{A}\dot{B}}{AB}-\dfrac{\ddot{C}}{C}-\dfrac{B^{\prime\prime}}{A^2B^2}
    +\left(\dfrac{\omega_z^{(de)}-\omega_y^{(de)}}{\omega_z^{(de)}-\omega_y^{(de)}}\right)\left[\dfrac{\ddot{A}}{A}-\dfrac{\dot{B}\dot{C}}{BC}+\dfrac{B'C'}{A^2BC}\right]\\
        \\
         \,\,\,\,\,\,\,\,\,\,\,\,\,\,\,\,\,\,\,\,\,\,\,\,\,\,\,\,\,\,\,\,\,\,\,\,\,\,\,\,
   +\left(\dfrac{\omega_x^{(de)}-\omega_z^{(de)}}{\omega_z^{(de)}-\omega_y^{(de)}}\right)\left[\dfrac{\ddot{B}}{B}-\dfrac{\dot{A}\dot{C}}{AC}+\dfrac{C''}{A^2C}\right]
   =0,
  \end{array}
%\right.
\end{equation}

\begin{equation}  \label{u210-2}
  %\left{
\begin{array}{ll}
   E_2=\dfrac{\dot{C}^{\prime}}{C}+\dfrac{\dot{B}^{\prime}}{B}-\dfrac{\dot{A}}{A}\left(\dfrac{C^{\prime}}{C}+\dfrac{B^{\prime}}{B}\right)=0,
  \end{array}
%\right.
\end{equation}
where
\begin{equation}  \label{u210-3}
  %\left{
\begin{array}{ll}
        p^{(pf)}(x,t)\,=\,\dfrac{\omega_x^{(de)}}{\omega_x^{(de)}-\omega_y^{(de)}}\left(\dfrac{C''}{A^2C}-\dfrac{\dot{A}\dot{C}}{AC}-\dfrac{\ddot{A}}{A}\right)
        -\dfrac{\omega_y^{(de)}}{\omega_x^{(de)}-\omega_y^{(de)}}\left(\dfrac{B'C'}{BC}-\dfrac{\dot{B}\dot{C}}{BC}-\dfrac{\ddot{B}}{B}\right)-\dfrac{\ddot{C}}{C},
  \end{array}
%\right.
\end{equation}

\begin{equation}  \label{u210-4}
  %\left{
\begin{array}{ll}
        \rho^{(pf)}(x,t)\,=\,\dfrac{1}{\omega_x^{(de)}-\omega_y^{(de)}}\left(\dfrac{\dot{B}\dot{C}}{BC}+\dfrac{\ddot{B}}{B}+\dfrac{C''}{A^2C}
        -\dfrac{\ddot{A}}{A}-\dfrac{\dot{A}\dot{C}}{AC}-\dfrac{B'C'}{A^2BC}\right)\\
        \\
         \,\,\,\,\,\,\,\,\,\,\,\,\,\,\,\,\,\,\,\,\,\,\,\,\,\,\,\,\,\,\,\,\,\,\,\,\,\,\,\,
+\dfrac{1}{A^2}\left(\dfrac{B'C'}{BC}+\dfrac{B''}{B}+\dfrac{C''}{A^C}\right)
        -\dfrac{\dot{A}\dot{B}}{AB}-\dfrac{\dot{A}\dot{C}}{AC}-\dfrac{\dot{B}\dot{C}}{BC},
  \end{array}
%\right.
\end{equation}

\begin{equation}  \label{u210-5}
  %\left{
\begin{array}{ll}
       \rho^{(de)}(x,t)\,=\,\,\dfrac{1}{\omega_x^{(de)}-\omega_y^{(de)}}\left(\dfrac{B'C'}{A^2BC}+\dfrac{\ddot{A}}{A}+\dfrac{\dot{A}\dot{C}}{AC}
        -\dfrac{\dot{B}\dot{C}}{BC}-\dfrac{\ddot{B}}{B}-\dfrac{C''}{A^2C}\right).
  \end{array}
%\right.
\end{equation}
%%%%%%%%%%%%%%%%%%%%%%%%%%%%%%%%%%%%%%%%%%%%%%%%%%%%%%%%
%%%%%%%%%%%%%%%%%%%%%%%%%%%%%%%%%%%%%%%%%%%%%%%%%%%%%%%%
\section{Lie point symmetry \& Similarity solutions}
%%%%%%%%%%%%%%%%%%%%%%%%%%%%%%%%%%%%%%%%%%%%%%%%%%%%%%%
Following Yadav and Ali \cite{anil2014}, we have acquired an
optimal system of one-dimensional subalgebras to be those spanned
by:
\begin{equation}\label{u32-5}
\begin{array}{ll}
\{X^{(1)}=X_1+a_3\,X_3+a_5\,X_5+a_6\,X_6,\,\,\,\,\,\,X^{(2)}=X_1+a_4\,X_4+a_5\,X_5+a_6\,X_6,\,\\
\,\,\,X^{(3)}=X_2+a_3\,X_3+a_5\,X_5+a_6\,X_6,\,\,\,\,\,\,X^{(4)}=X_2+a_4\,X_4+a_5\,X_5+a_6\,X_6,\,\\
\,\,\,X^{(5)}=X_3+a_5\,X_5+a_6\,X_6,\,\,\,\,\,X^{(6)}=X_4+a_5\,X_5+a_6\,X_6,\,\,\,\,\,
X^{(7)}=X_5+a_6\,X_6,\,\,\,\,\,X^{(8)}=X_6\}.
\end{array}
\end{equation}

If we considered the symmetries $X^{(5)}$ or $X^{(6)}$ or $X^{(7)}$ or $X^{(8)}$, then $a_1=a_2=0$, 
we shall analyze the similarity solutions associated with the optimal systems of 
symmetries $X^{(1)}$, $X^{(2)}$, $X^{(3)}$ and $X^{(4)}$ only as the following:\\

\textbf{Solution (I):} The symmetries $X^{(1)}$ has the characteristic equations:
\begin{equation}\label{u41-1}
\dfrac{dx}{x}=\dfrac{dt}{a_3\,t}=\dfrac{dB}{a_5\,B}=\dfrac{dC}{a_6\,C}.
\end{equation}
Then the similarity variable and the similarity transformations takes the form:
\begin{equation}\label{u42-1}
\begin{array}{ll}
\xi=\dfrac{x^a}{t},\,\,\,\,\,\,B(x,t)=\,x^{b}\,\Psi(\xi),\,\,\,\,\,\,C(x,t)=\,x^{c}\,\Phi(\xi),
\end{array}
\end{equation}
where $a=a_3$, $b=a_5$ and $c=a_6$ are an arbitrary constants. In
this case, we have

\begin{equation}\label{u42-2}
%\left\{
  \begin{array}{ll}
    A(t)=d\,t^{1-\frac{1}{a}},   \,\,\,\,\,\,\,
    \omega_y^{(de)}(t)=\omega_x^{(de)}(t)+q\,t^{\frac{2}{a}-2},
\,\,\,\,\,\,\,
    \omega_z^{(de)}(t)=\omega_x^{(de)}(t)+r\,t^{\frac{2}{a}-2},
  \end{array}
%\right.
\end{equation}
where $d=a_7\,a_3^{1-\frac{1}{a}}$, $q=a_8\,a_3^{\frac{2}{a_3}-2}$
and $r=a_9\,a_3^{\frac{2}{a_3}-2}$ are an arbitrary constants.\\
In conection with our earlier work \cite{anil2014}, we can find the solution of the
Einstein field equations as the following:\\

\begin{equation}\label{uu1}
\left\{
  \begin{array}{ll}
A(t)=d\,t^{1+\frac{\beta_4}{2}},\,\,\,\,\,
B(x,t)=\beta_1\,x^{-\frac{(2+\beta_4)(2+3\,\beta_4)}{2\,\beta_4\,(1+\beta_4)}}\,t^{1+\frac{2}{\beta_4}},\,\,\,\,\,
C(x,t)=\beta_3\,x^{1+\frac{2}{\beta_4}}\,\Big(t^{-\beta_4}-d_0^2\,x^2\Big),\\
\\
\omega_y(t)=\omega_x(t)+q\,t^{-2-\beta_4},\,\,\,\,\,\,\,\,\omega_z(t)=\omega_x(t)+r\,t^{-2-\beta_4},\,\,\,\,\,\,\,\,
d^2=\dfrac{d_0^2\,(64+272\,\beta_4+408\,\beta_4^2+256\,\beta_4^3+57\,\beta_4^4)}{4\,\beta_4^3\,(1+\beta_4)^3},
  \end{array}
\right.
\end{equation}

where $d_0$, $q$, $\beta_1$, $\beta_2$, $\beta_3$ and $\beta_4$ are an arbitrary constants, while
$\omega_x$ is an arbitrary function of $t$. It is observed from
equations (\ref{uu1}), the line element (\ref{spacetime}) can be
written in the following form:\\

\begin{equation}  \label{s1}
\begin{array}{ll}
ds_{1}^2=d^2\,t^{2+\beta_4}\,dx^2+\beta_1^2\,x^{-\frac{(2+\beta_4)(2+3\,\beta_4)}{\beta_4\,(1+\beta_4)}}\,t^{2+\frac{4}{\beta_4}}\,dy^2
+\beta_3^2\,x^{2+\frac{4}{\beta_4}}\,\Big(t^{-\beta_4}-d_0^2\,x^2\Big)^2\,dz^2-dt^2.
\end{array}
\end{equation}

\textbf{Remark:} In the above solution, we can replace $t$ by $t+\delta_1$ and $x$ by $x+\delta_2$ without loss of generality, where $\delta_1$ and $\delta_2$ are an some arbitrary constants.\\
%%%%%%%%%%%%%%%%%%%%%%%%%%%%%%%%%%%%%%%%%%%%%%%%%%%%%%%%%%%%%%%%%%%%%%%
\textbf{Solution (II):} The symmetries $X^{(3)}$ has the
characteristic equations:
\begin{equation}\label{u61-1}
\dfrac{dx}{1}=\dfrac{dt}{a_3\,t}=\dfrac{dB}{a_5\,B}=\dfrac{dC}{a_6\,C}.
\end{equation}
Then the similarity variable and the similarity transformations
takes the form:
\begin{equation}\label{u62-1}
\begin{array}{ll}
\xi=t\,\exp\big[a\,x\big],\,\,\,\,\,\,B(x,t)=\Psi(\xi)\,\exp\big[b\,x\big],\,\,\,\,\,\,C(x,t)=\Phi(\xi)\,\exp\big[c\,x\big],
\end{array}
\end{equation}
where $a=-\frac{1}{a_3}$, $b=a_5$ and $c=a_6$ are an arbitrary
constants. In this case, we have
\begin{equation}\label{u62-2}
%\left\{
  \begin{array}{ll}
    A(t)=d\,t,\,\,\,\,\,\,\,
    \omega_y^{(de)}(t)=\omega_x^{(de)}(t)+\frac{q}{t^2},\,\,\,\,\,\,\,
    \omega_z^{(de)}(t)=\omega_x^{(de)}(t)+\frac{r}{t^2},
  \end{array}
%\right.
\end{equation}
where $d=a_7$, $q=a_8$ and $r=a_9$ are an arbitrary constants.\\
Finally in light of ref. \cite{anil2014}, one can find the solution of the Einstein field equations as the
following:

\begin{equation}\label{uu1}
\left\{
  \begin{array}{ll}
A(t)=a\,t,\,\,\,\,\,\,\,\,\omega_y^{(de)}(t)=\omega_x^{(de)}(t)+\dfrac{q}{t^2},\,\,\,\,\,\,\,\,
    \omega_z^{(de)}(t)=\omega_x^{(de)}(t),\\
\\
B(x,t)=\beta_4\,t^{-b/a}\,\left[\beta_3+\Big(t\,e^{a\,x}\Big)^{\frac{b^2+2\,a\,c+c^2}{2\,a\,b}}\right]^{\frac{b\,(b+c)}{b^2+c^2}},\\
\\
C(x,t)=\beta_2\,t\,e^{(a+c)\,x}\,\left[\beta_3+\Big(t\,e^{a\,x}\Big)^{\frac{b^2+2\,a\,c+c^2}{2\,a\,b}}\right]^{\frac{b\,(b-c)}{b^2+c^2}},\\
\end{array}
\right.
\end{equation}
where $a$, $b$ $c$, $\beta_2$, $\beta_3$, $\beta_4$ and $q$ are an
arbitrary constants, while $\omega_x$ is an arbitrary function of
$t$. It is observed from equations (\ref{uu1}), the line element
(\ref{spacetime}) can be written in the following form:
\begin{equation}  \label{s2}
\begin{array}{ll}
ds_{2}^2=a^2\,t^2\,dx^2+\beta_4^2\,t^{-2\,b/a}\,\left[\beta_3+\Big(t\,e^{a\,x}\Big)^{\frac{b^2+2\,a\,c+c^2}{2\,a\,b}}\right]^{\frac{2\,b\,(b+c)}{b^2+c^2}}\,dy^2\\
\\
\,\,\,\,\,\,\,\,\,\,\,\,\,\,\,\,\,\,\,\,\,\,\,\,\,\,\,\,\,\,\,\,\,\,\,\,\,\,\,\,
+\beta_2^2\,t^2\,e^{2\,(a+c)\,x}\,\left[\beta_3+\Big(t\,e^{a\,x}\Big)^{\frac{b^2+2\,a\,c+c^2}{2\,a\,b}}\right]^{\frac{2\,b\,(b-c)}{b^2+c^2}}\,dz^2-dt^2.
\end{array}
\end{equation}

%%%%%%%%%%%%%%%%%%%%%%%%%%%%%%%%%%%%%%%%%%%%%%%%%%%%%%%%
\section{Physical aspects of the models}
%%%%%%%%%%%%%%%%%%%%%%%%%%%%%%%%%%%%%%%%%%%%%%%%%%%%%%%

\textbf{For the Model (\ref{s1}):}\\

The expressions of $p^{(pf)}$, $\rho^{(pf)}$ and $\rho^{(de)}$  for
the model (\ref{s1}), are given by:

\begin{equation}\label{uu1-1}
%\left\{
  \begin{array}{ll}
p^{(pf)}(x,t)=\dfrac{\beta_4}{4\,f(x,t)}\,\Bigg[
8\,q\,(1+\beta_4)^2\,(2+\beta_4)^2\,(2+3\,\beta_4)\,t^{-2-\beta_4}\\
\,\,\,\,\,\,\,\,\,\,\,\,\,\,\,\,\,\,\,\,\,\,\,\,\,\,\,\,\,\,\,\,\,\,\,\,\,\,\,\,\,\,\,
+q\,d_0^2\,\Big[320+1728\,\beta_4+3600\,\beta_4^2+3616\,\beta_4^3+1750\,\beta_4^4+327\,\beta_4^5\Big]\,x^2\,t^{-2}\\
\,\,\,\,\,\,\,\,\,\,\,\,\,\,\,\,\,\,\,\,\,\,\,\,\,\,\,\,\,\,\,\,\,\,\,\,\,\,\,\,\,\,\,
-q\,d_0^4\,(2+\beta_4)\,(4+3\,\beta_4)\,\Big[16+56\,\beta_4+60\,\beta_4^2+19\,\beta_4^3\Big]\,x^2\,t^{-2+\beta_4}\\
\,\,\,\,\,\,\,\,\,\,\,\,\,\,\,\,\,\,\,\,\,\,\,\,\,\,\,\,\,\,\,\,
+4\,(1+\beta_4)\,(8+12\,\beta_4+3\,\beta_4^2)\Big[2\,(1+\beta_4)\,(2+\beta_4)+d_0^2\,(4+10\,\beta_4+5\,\beta_4^2)\,x^2\,t^{\beta_4}\Big]\omega_x^{(pf)}(t)
\Bigg],
  \end{array}
%\right.
\end{equation}

\begin{equation}\label{uu1-2}
%\left\{
  \begin{array}{ll}
\rho^{(pf)}(x,t)=\dfrac{\beta_4}{4\,f(x,t)}\,\Bigg[
4\,(1+\beta_4)\,(8+12\,\beta_4+3\,\beta_4^2)\Big[2\,(1+\beta_4)\,(2+\beta_4)+d_0^2\,(4+10\,\beta_4+5\,\beta_4^2)\,x^2\,t^{\beta_4}\Big]\\
\,\,\,\,\,\,\,\,\,\,\,\,\,\,\,\,\,\,\,\,\,\,\,\,\,\,\,\,\,\,\,\,\,\,\,\,\,\,\,\,\,\,\,\,\,\,\,\,\,\,\,\,\,
-4\,q\,(1+\beta_4)\,(2+\beta_4)\,(8+24\,\beta_4+26\,\beta_4^2+9\,\beta_4^3)\,t^{-2-\beta_4}\\
\,\,\,\,\,\,\,\,\,\,\,\,\,\,\,\,\,\,\,\,\,\,\,\,\,\,\,\,\,\,\,\,\,\,\,\,\,\,\,\,\,\,\,\,\,\,\,\,\,\,\,\,\,
-q\,d_0^2\,\Big[320+1344\,\beta_4+1968\,\beta_4^2+1160\,\beta_4^3+194\,\beta_4^4-27\,\beta_4^5\Big]\,x^2\,t^{-2}\\
\,\,\,\,\,\,\,\,\,\,\,\,\,\,\,\,\,\,\,\,\,\,\,\,\,\,\,\,\,\,\,\,\,\,\,\,\,\,\,\,\,\,\,\,\,\,\,\,\,\,\,\,\,
+q\,d_0^4\,(2+\beta_4)\,(4+3\,\beta_4)\,\Big[16+56\,\beta_4+60\,\beta_4^2+19\,\beta_4^3\Big]\,x^2\,t^{-2+\beta_4}

\Bigg],
  \end{array}
%\right.
\end{equation}

\begin{equation}\label{uu1-3}
%\left\{
  \begin{array}{ll}
\rho^{(de)}(x,t)=-\dfrac{\beta_4\,(1+\beta_4)\,(8+12\,\beta_4^2+3\,\beta_4^2)\,\Big[
2\,(1+\beta_4)\,(2+\beta_4)+d_0^2\,(4+10\,\beta_4+5\,\beta4)\,x^2\,t^{\beta_4}\Big]}{f(x,t)},
  \end{array}
%\right.
\end{equation}
where
$$
f(x,t)=q\,d_0^2\,(4+3\,\beta_4)\,\Big[16+56\,\beta_4+60\,\beta_4^2+19\,\beta_4^3\Big]\Big(d_0^2\,x^4\,t^{\beta_4}-x^2\Big).
$$
The volume element is
\begin{equation}  \label{uu1-4}
V=d\,\beta_1\,\beta_3\,x^{-\frac{2+\beta_4}{2+2\,\beta_4}}\,t^{1-\beta_4}\Big(d_0^2\,x^2\,t^{\beta_4}-1\Big).
\end{equation}
The expansion scalar, which determines the volume behavior of the fluid, is given by:
\begin{equation}\label{uu1-5}
%\left\{
  \begin{array}{ll}
\Theta=\dfrac{1}{t}\,\Big(1+\dfrac{\beta_4}{d_0^2\,x^2\,t^{\beta_4}-1}\Big).
  \end{array}
%\right.
\end{equation}
The non-vanishing components of the shear tensor, $\sigma_i^j$, are:
\begin{equation}\label{uu1-6}
%\left\{
  \begin{array}{ll}
\dfrac{\sigma_1^1}{\Theta}\,=\,\dfrac{d_0^2\,(4+3\,\beta_4)\,x^2\,t^{\beta_4}-4-5\,\beta_4}{6\,\Big[\beta_4-1+d_0^2\,x^2\,t^{\beta_4}\Big]},
  \end{array}
%\right.
\end{equation}

\begin{equation}\label{uu1-7}
%\left\{
  \begin{array}{ll}
\dfrac{\sigma_2^2}{\Theta}\,=\,\dfrac{2+\beta_4-d_0^2\,(4+3\,\beta_4)\,x^2\,t^{\beta_4}}{6\,\Big[\beta_4-1+d_0^2\,x^2\,t^{\beta_4}\Big]},
  \end{array}
%\right.
\end{equation}

\begin{equation}\label{uu1-8}
%\left\{
  \begin{array}{ll}
\dfrac{\sigma_3^3}{\Theta}\,=\,\dfrac{1+2\,\beta_4-d_0^2\,x^2\,t^{\beta_4}-4-5\,\beta_4}{3\,\Big[\beta_4-1+d_0^2\,x^2\,t^{\beta_4}\Big]}.
  \end{array}
%\right.
\end{equation}

The shear scalar is:
\begin{equation}\label{uu1-9}
%\left\{
  \begin{array}{ll}
\dfrac{\sigma^2}{\Theta^2}\,=\,\dfrac{4+10\,\beta_4+7\,\beta_4^2-2\,d_0^2\,(4+8\,\beta_4+3\,\beta_4^2)\,x^2\,t^{\beta_4}
+d_0^4\,(4+6\,\beta_4+3\,\beta_4^2)\,x^4\,t^{2\,\beta_4}}{12\,\Big[\beta_4-1+d_0^2\,x^2\,t^{\beta_4}\Big]^2}.
  \end{array}
%\right.
\end{equation}

The deceleration parameter is given by \cite{fein1, rayc1}
\begin{equation}\label{uu1-11}
%\left\{
  \begin{array}{ll}
\mathbf{q}=\dfrac{\Big(\beta_4-1+d_0^2\,x^2\,t^{\beta_4}\Big)\Big[(1-\beta_4)\,(2+\beta_4)-d_0^2\,(4-\beta_4-3\,\beta_4^2)\,x^2\,t^{\beta_4}
+2\,d_0^4\,x^4\,t^{2\,\beta_4}\Big]}{t^4\,\Big[d_0^2\,x^2\,t^{\beta_4}-1\Big]^4}.
  \end{array}
%\right.
\end{equation}

%%%%%%%%%%%%%%%%%%%%%%%%%%%%%%%%%%%%%%%%%%%%%%%%%%%%%%%%%%%%%%%%%%%%%%%%%%%%
\begin{figure*}[thbp]
\begin{tabular}{rl}
\includegraphics[width=8cm]{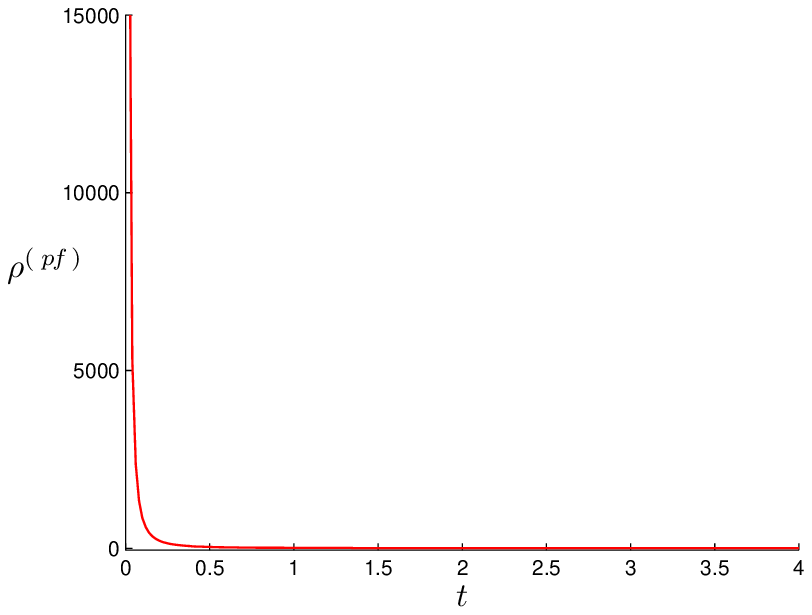}
\includegraphics[width=8cm]{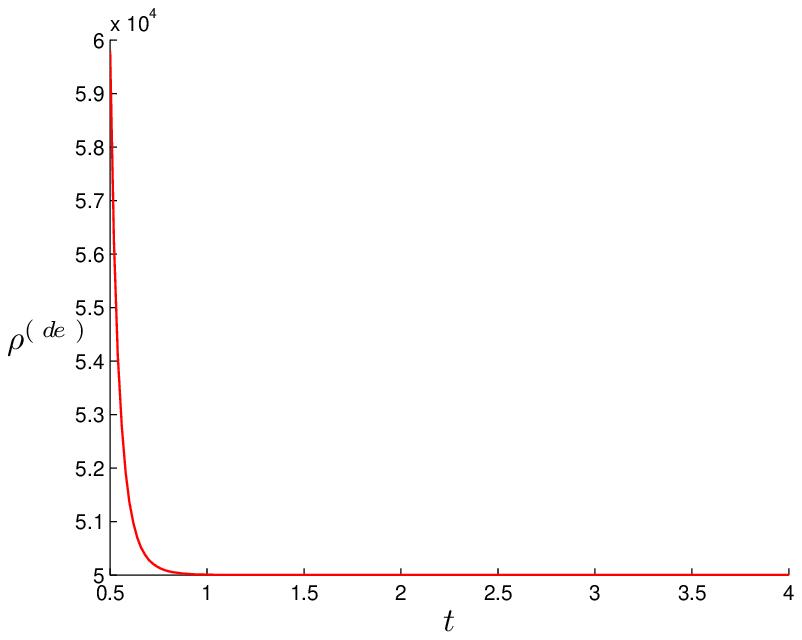}\\
\includegraphics[width=8cm]{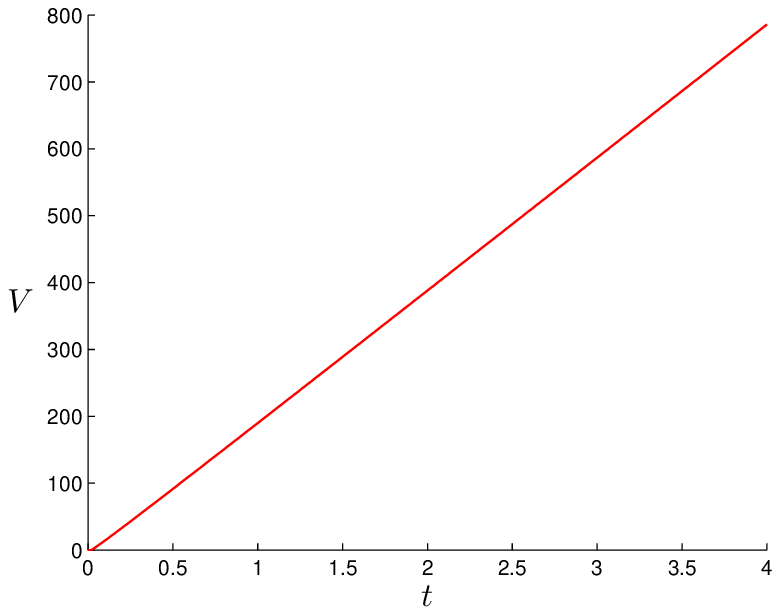}
\includegraphics[width=8cm]{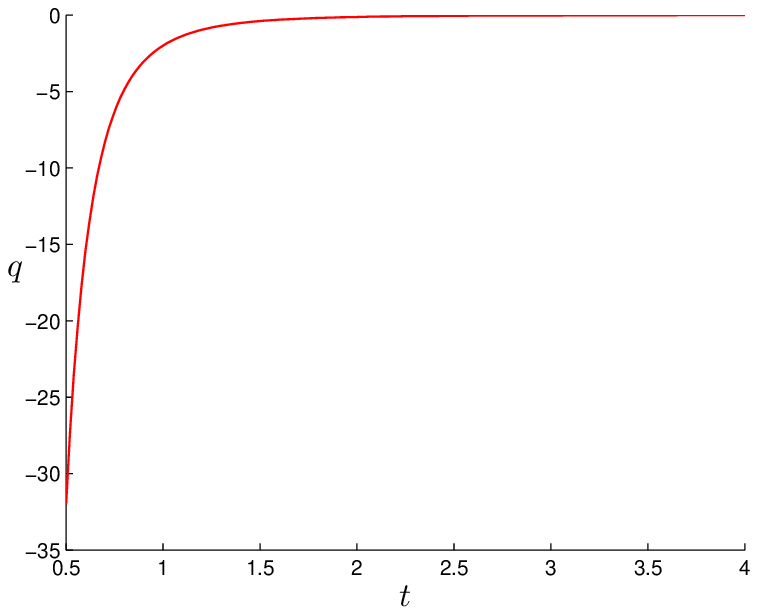}
\end{tabular}
\caption{Variation of energy density (upper left panel), dark energy density (upper right panel),
volume (lower left panel) and DP (lower right panel) versus time for model (\ref{s1})}
\end{figure*}
%%%%%%%%%%%%%%%%%%%%%%%%%%%%%%%%%%%%%%%%%%%%%%%%%%%%%%%%%%%%%%%%%%%%%%%
%%%%%%%%%%%%%%%%%%%%%%%%%%%%%%%%%%%%%%%%%%%%%%%%%%%%%%%%%%%%%%%%%%%%%%%%%%%%
\begin{figure*}[thbp]
\begin{tabular}{rl}
\includegraphics[width=8cm]{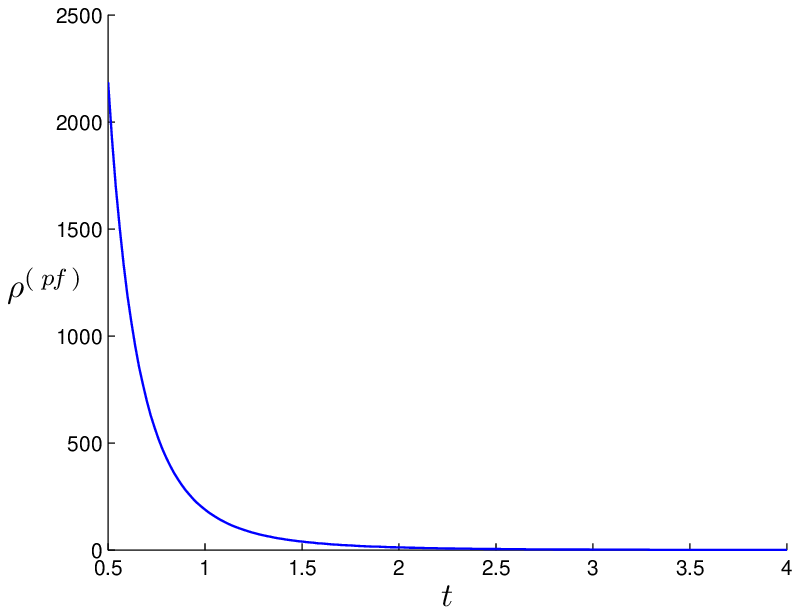}
\includegraphics[width=8cm]{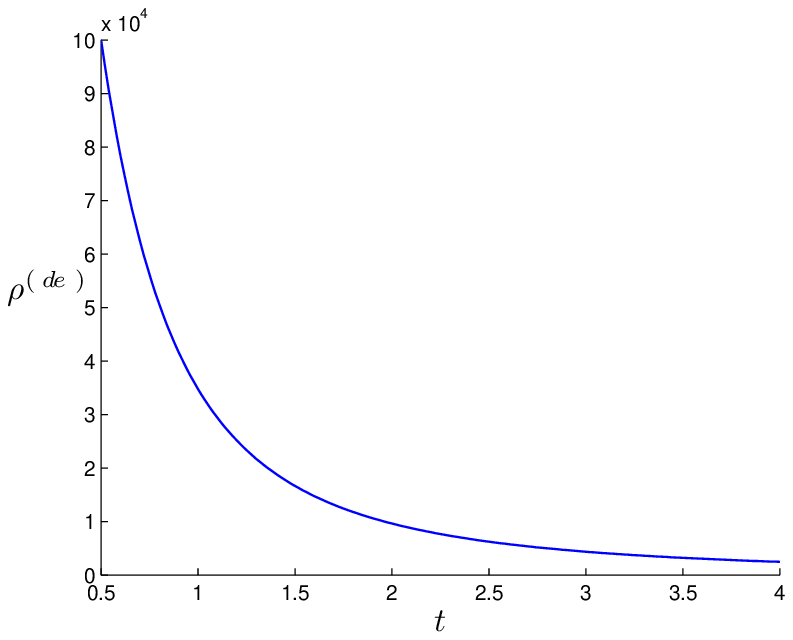}\\
\includegraphics[width=8cm]{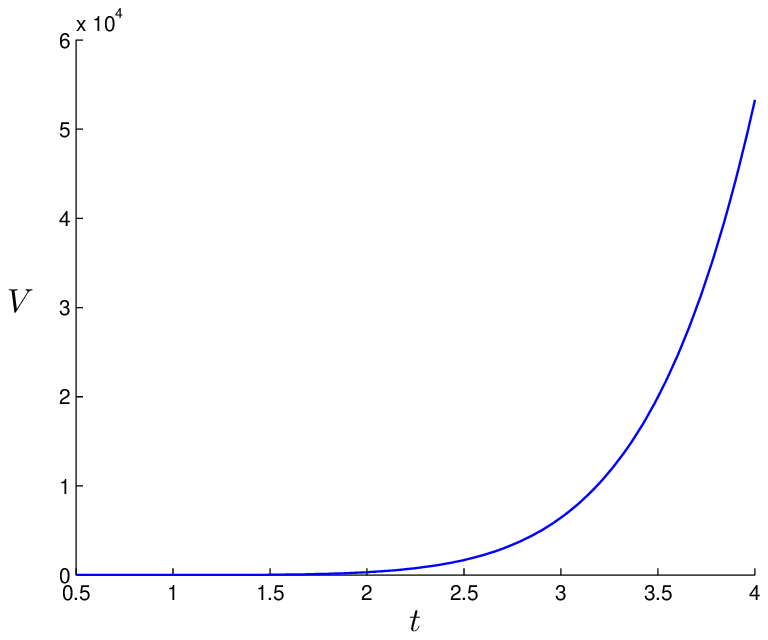}
\includegraphics[width=8cm]{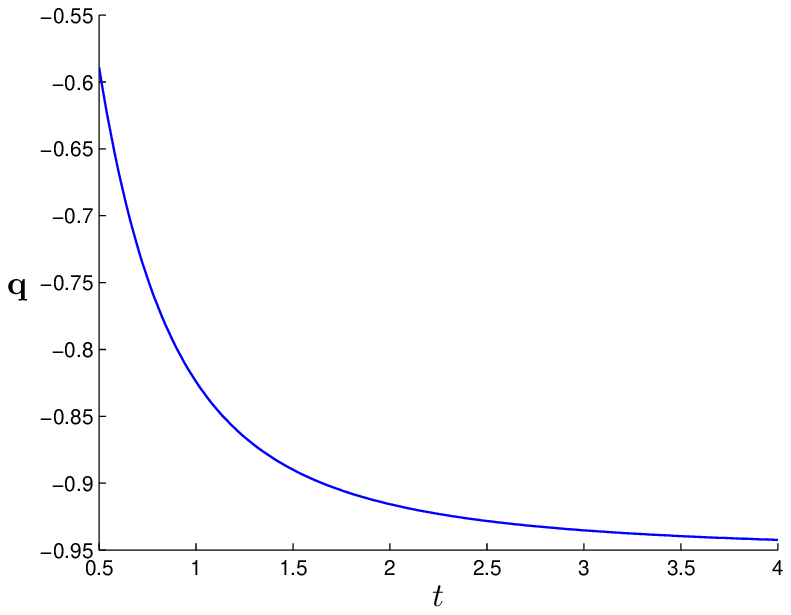}
\end{tabular}
\caption{Variation of energy density (upper left panel), dark energy density (upper right panel),
volume (lower left panel) and DP (lower right panel) versus time for model (\ref{s2})}
\end{figure*}
%%%%%%%%%%%%%%%%%%%%%%%%%%%%%%%%%%%%%%%%%%%%%%%%%%%%%%%%%%%%%%%%%%%%%%%

\textbf{For the Model (\ref{s2}):}\\

The expressions of $p^{(pf)}$, $\rho^{(pf)}$ and $\rho^{(de)}$  for
the model (\ref{s2}), are given by:

\begin{equation}\label{uu2-1}
%\left\{
  \begin{array}{ll}
p^{(pf)}(x,t)=\dfrac{b\,c\,q\,\big[b^2+c^2+2\,a\,(b+c)\big]\Big(t\,e^{a\,x}\Big)^{\frac{b^2+2\,a\,c+c^2}{2\,a\,b}}
-\beta_3\,(b^2+c^2)\,\Big[b^2\,q+(b^2+2\,a\,c+c^2)\,t^2\,\omega_x^{(de)}(t)\Big]}{g(x,t)},
  \end{array}
%\right.
\end{equation}

\begin{equation}\label{uu2-2}
%\left\{
  \begin{array}{ll}
\rho^{(pf)}(x,t)=\dfrac{1}{g(x,t)}\,\Bigg(q\,(b^2+b\,c+c^2)\,\big[b^2+c^2+2\,a\,(b+c)\big]\,\Big(t\,e^{a\,x}\Big)^{\frac{b^2+2\,a\,c+c^2}{2\,a\,b}}\\
\\
\,\,\,\,\,\,\,\,\,\,\,\,\,\,\,\,\,\,\,\,\,\,\,\,\,\,\,\,\,\,\,\,\,\,\,\,\,\,\,\,\,\,\,\,\,\,\,\,\,\,\,\,\,\,\,\,\,\,\,\,
\,\,\,\,\,\,\,\,\,\,\,\,\,\,\,\,\,\,\,\,
+\beta_3\,(b^2+c^2)\,\Big[q\,\big[c^2+2\,a\,(b+c)\big]-(b^2+2\,a\,c+c^2)\,t^2\Big]\Bigg),
  \end{array}
%\right.
\end{equation}
\begin{equation}\label{uu2-3}
%\left\{
  \begin{array}{ll}
\rho^{(de)}(x,t)=\dfrac{\beta_3\,(b^2+c^2)\,(b^2+2\,a\,c+c^2)\,t^2}{g(x,t)},
  \end{array}
%\right.
\end{equation}
where
$g(x,t)=a^2\,q\,(b^2+c^2)\,t^2\,\Big[\beta_3+\Big(t\,e^{a\,x}\Big)^{\frac{b^2+2\,a\,c+c^2}{2\,a\,b}}\Big]$.\\
The volume element is
\begin{equation}  \label{uu2-4}
V=a\,\beta_2\,\beta_4\,t^{2-b/a}\,e^{(a+c)\,x}\,\Big[\beta_3+\Big(t\,e^{a\,x}\Big)^{\frac{b^2+2\,a\,c+c^2}{2\,a\,b}}\Big]^{\frac{2\,b^2}{b^2+c^2}}.
\end{equation}
The expansion scalar, which determines the volume behavior of the fluid, is given by:
\begin{equation}\label{uu2-5}
%\left\{
  \begin{array}{ll}
\Theta=\dfrac{1}{(a^2+c^2)\,t}\Bigg(2\,(b^2+b\,c+c^2)-\dfrac{b\,\beta_3\,(b^2+2\,a\,c+c^2)}{a\,\Big[\beta_3+\Big(t\,e^{a\,x}\Big)^{\frac{b^2+2\,a\,c+c^2}{2\,a\,b}}\Big]}\Bigg),
  \end{array}
%\right.
\end{equation}
The non-vanishing components of the shear tensor, $\sigma_i^j$, are:
\begin{equation}\label{uu2-6}
%\left\{
  \begin{array}{ll}
\dfrac{\sigma_1^1}{\Theta}\,=\,\dfrac{\beta_3\,(a+b)\,(b^2+c^2)+a\,(b-c)^2\,\Big(t\,e^{a\,x}\Big)^{\frac{b^2+2\,a\,c+c^2}{2\,a\,b}}}{
3\,\beta_3\,(2\,a-b)\,(b^2+c^2)+6\,a\,(b^2+b\,c+c^2)^2\,\Big(t\,e^{a\,x}\Big)^{\frac{b^2+2\,a\,c+c^2}{2\,a\,b}}},
  \end{array}
%\right.
\end{equation}

\begin{equation}\label{uu2-7}
%\left\{
  \begin{array}{ll}
\dfrac{\sigma_2^2}{\Theta}\,=\,\dfrac{4\,\beta_3\,(a+b)\,(b^2+c^2)+(b-c)\,\big[2\,a\,(2\,b+c)+3\,(b^2+c^2)\big]\,\Big(t\,e^{a\,x}\Big)^{\frac{b^2+2\,a\,c+c^2}{2\,a\,b}}}{
6\,\beta_3\,(b-2\,a)\,(b^2+c^2)-12\,a\,(b^2+b\,c+c^2)^2\,\Big(t\,e^{a\,x}\Big)^{\frac{b^2+2\,a\,c+c^2}{2\,a\,b}}},
  \end{array}
%\right.
\end{equation}

\begin{equation}\label{uu2-8}
%\left\{
  \begin{array}{ll}
\dfrac{\sigma_3^3}{\Theta}\,=\,\dfrac{2\,\beta_3\,(a+b)\,(b^2+c^2)+(b-c)\,\big[2\,a\,(b+2\,c)+3\,(b^2+c^2)\big]\,\Big(t\,e^{a\,x}\Big)^{\frac{b^2+2\,a\,c+c^2}{2\,a\,b}}}{
6\,\beta_3\,(2\,a-b)\,(b^2+c^2)+12\,a\,(b^2+b\,c+c^2)^2\,\Big(t\,e^{a\,x}\Big)^{\frac{b^2+2\,a\,c+c^2}{2\,a\,b}}}.
  \end{array}
%\right.
\end{equation}

The shear scalar is:
\begin{equation}\label{uu2-9}
%\left\{
  \begin{array}{ll}
\dfrac{\sigma^2}{\Theta^2}\,=\,\dfrac{1}{h(x,t)}\,\Bigg(
4\,\beta_3^2\,(a+b)^2\,(b^2+c^2)^2\\
\\
\,\,\,\,\,\,\,\,\,\,\,\,\,\,\,\,\,\,\,\,\,\,\,\,\,\,\,\,\,\,\,\,\,\,\,\,\,\,\,\,
+2\beta_3\,(a+b)\,(b-c)\,(b^2+c^2)\,\big[2\,a\,(2\,b+c)+3\,(b^2+c^2)\big]\,\Big(t\,e^{a\,x}\Big)^{\frac{b^2+2\,a\,c+c^2}{2\,a\,b}}\\
\\
\,\,\,\,\,\,\,\,\,\,\,\,\,\,\,\,\,\,\,\,\,\,\,\,\,\,\,\,\,\,\,\,\,\,\,\,\,\,\,\,
+(b-c)^2\,\big[6\,a\,(b+c)\,(b^2+c^2)+3\,(b^2+c^2)^2+4\,a\,(b^2+b\,c
+c^2)\big]\,\Big(t\,e^{a\,x}\Big)^{\frac{b^2+2\,a\,c+c^2}{a\,b}}\Bigg),
  \end{array}
%\right.
\end{equation}
where
\begin{equation}\label{uu2-9-2}
%\left\{
  \begin{array}{ll}
h(x,t)=12\Bigg[\beta_3\,(b-2\,a)\,(b^2+c^2)-2\,a\,(b^2+b\,c+c^2)^2\,\Big(t\,e^{a\,x}\Big)^{\frac{b^2+2\,a\,c+c^2}{2\,a\,b}}\Bigg]^2.
  \end{array}
%\right.
\end{equation}

The deceleration parameter is given by:
\begin{equation}\label{uu2-11}
%\left\{
  \begin{array}{ll}
\mathbf{q}=\dfrac{1}{k(x,t)}\,\Bigg(
2\,\beta_3^2\,(a+b)\,(2\,a-b)\,(b^2+c^2)^2-\beta_3\,(b^2+c^2)\,\Big[3\,(b^2+c^2)^2\\
\\
\,\,\,\,\,\,\,\,\,\,\,\,\,\,\,\,\,\,\,\,\,\,\,\,\,\,\,\,\,\,\,\,\,\,\,\,\,\,\,\,
+4\,a^2\,(c^2+b\,c-2\,b^2)-2\,a\,(b^3-2\,b^2\,c+b\,c^2-6\,c^3)\Big]\,\Big(t\,e^{a\,x}\Big)^{\frac{b^2+2\,a\,c+c^2}{2\,a\,b}}\\
\\
\,\,\,\,\,\,\,\,\,\,\,\,\,\,\,\,\,\,\,\,\,\,\,\,\,\,\,\,\,\,\,\,\,\,\,\,\,\,\,\,
+4\,a^2(b-c)^2\,(b^2+b\,c+c^2)\,\Big(t\,e^{a\,x}\Big)^{\frac{b^2+2\,a\,c+c^2}{a\,b}}\Bigg)\\
\,\,\,\,\,\,\,\,\,\,\,\,\,\,\,\,\,\,\,\,\,\,\,\,\,\,\,\,\,\,\,\,\,\,\,\,\,\,\,\,
\times\,\Bigg(b\,\beta_3\,(b^2+2\,a\,c+c^2)+2\,a\,(b^2+b\,c+c^2)^2\,\Big(t\,e^{a\,x}\Big)^{\frac{b^2+2\,a\,c+c^2}{2\,a\,b}}\Bigg)^2,
  \end{array}
%\right.
\end{equation}
where
\begin{equation}\label{uu2-11-2}
%\left\{
  \begin{array}{ll}
k(x,t)=2\,a^4\,(b^2+c^2)^4\,t^4\,\Bigg[\beta_3+\Big(t\,e^{a\,x}\Big)^{\frac{b^2+2\,a\,c+c^2}{2\,a\,b}}\Bigg]^4.
  \end{array}
%\right.
\end{equation}

\section {Conclusion}
In this paper, we have investigated the role of DE with time dependent skewness
parameters along the spatial directions that quantify the deviation of pressure
from isotropy. Generally, the models behave like an expanding, shearing and non-rotating model of universe. Here, we obtain two similarity dark energy models (\ref{s1}) and (\ref{s2}) associated with $X^{(1)}$ and $X^{(3)}$ respectively. The main features of the work are as follows:
\begin{itemize}
 \item The models are based on similarity solution of field equation and we have obtained the new class of
exact solution.
\item We have, in general discussed several physical features and geometrical
properties of the models. However, as a special case, most notable aspect of the solutions
have been studied that are non-singular in nature. All figures depicts interesting
features of the present cosmological models in terms of DP and other physical parameters.
\item In the derived models, the matter energy density and dark energy density remains positive.
Therefore, the WEC and NEC are satisfied, which in turn imply that the derived models are
physically realistic.
\item As $t \rightarrow \infty$, $\rho^{(de)} \rightarrow 0$, hence the DE density is decreasing function of time.
\item The derived models seem to describes the dynamics of universe from big bang to present epoch and
DE dominates the universe at present time which may be attributed to the current accelerated expansion of
universe.
\item Hypothetical DE is the most reasonable way of explaining why the universe is expanding at an ever
increasing rate. DE plays a massive part in shaping our reality, however, it is to be noted that 
no body seems certain of what the dang
stuff actually is. Future space mission hope to solve this mystery and shake up our current understanding
of the universe. To our knowledge, this work is the first study of minimally interacting DE with normal matter in 
Bianchi - I inhomogeneous space-time and its general form.
\end{itemize}

\textbf{Acknowledgment:} This paper is funded by the Deanship of
Scientific Research (DSR), King Abdulaziz University, Jeddah, under
grant No. (130--682--D1435). The authors, therefore, acknowledge
with thank DSR technical and financial support.

%%%%%%%%%%%%%%%%%%%%%%%%%%%%%%%%%%%%%%%

\end{document}